# Approximate Ripple Carry and Carry Lookahead Adders – A Comparative Analysis

P. Balasubramanian, C. Dang, D.L. Maskell, and K. Prasad

*Abstract* - Approximate ripple carry adders (RCAs) and carry lookahead adders (CLAs) are presented which are compared with accurate RCAs and CLAs for performing a 32-bit addition. The accurate and approximate RCAs and CLAs are implemented using a 32/28nm CMOS process. Approximations ranging from 4- to 20-bits are considered for the less significant adder bit positions. The simulation results show that approximate RCAs report reductions in the power-delay product (PDP) ranging from 19.5% to 82% than the accurate RCA for approximation sizes varying from 4- to 20-bits. Also, approximate CLAs report reductions in PDP ranging from 16.7% to 74.2% than the accurate CLA for approximation sizes varying from 4- to 20-bits. On average, for the approximation sizes considered, it is observed that approximate CLAs achieve a 46.5% reduction in PDP compared to the approximate RCAs. Hence, approximate CLAs are preferable over approximate RCAs for the low power implementation of approximate computer arithmetic.

## I. INTRODUCTION

In practical applications such as digital signal processing, digital communication, machine learning, computer graphics, computer vision, data analytics, data mining, cloud computing, biometrics, neuromorphic computing etc. [1 – 3] golden results i.e. accurate computation results may not always be necessary. Instead, approximately correct computation results which are confined within a specified error bound may be acceptable. This is by taking advantage of the limitation of the human perception. The main driver behind approximate computing [4] is that many practical applications such as those mentioned above are inherently error-resilient. As a result, approximation could pave the way for utilizing less resources to produce an acceptable result rather than the usage of entire resources to produce the accurate result. The usage of less resources to perform approximate computing instead of the usage of ample resources to perform accurate computing translates into low power and energy efficiency [5]. Given the emergence of sophisticated technologies such as the Internet-of-Things, approximate computing may likely become the norm for future consumer electronics and mobile applications [6] [7] with a likelihood of accurate computing being reserved for electronic circuits and systems deployed in mission-critical and safety-critical applications.

Approximate computing can be broadly divided into three categories as approximate circuits [5], approximate storage [8] (both these can be grouped as hardware-level approximation), and software-level approximation [9]. With respect to approximate circuits, approximate synthesis of logic circuits [10], and arithmetic circuits i.e. adders and multipliers [5] are being given attention. With respect to approximate adders, static realizations [11 – 14] which incorporate a fixed degree of approximation and dynamic realizations [15 – 18] which incorporate varying degrees of approximation have been proposed. The static and dynamic approximate adder implementations either correspond to the gate level or the transistor level. Static implementations are application-specific and are better suited for ASIC designs. Dynamic implementations are rather flexible compared to static implementations since the approximation size may be varied subject to demand, and hence they are said to be application-generic. Dynamic implementations are ideally suited for FPGA based designs. Also, the dynamic implementations may be configured to produce accurate or approximate outputs based on demand. Nevertheless, the area and design complexities of dynamic implementations would be greater compared to the static implementations and also multiple clock cycles may be required to produce the accurate output which might result in greater power dissipation and hence are likely to reduce the throughput.

This paper presents approximate RCA and CLA structures which are compared with the accurate RCA and CLA to perform a 32-bit addition. The accurate and approximate RCAs and CLAs are implemented in semi-custom ASIC design style using a 32/28nm CMOS process. Section II briefly discusses the accurate RCA and CLA realizations. Section III discusses the principle of approximation, and then presents the approximate RCAs and CLAs. Section IV presents the design metrics estimated for accurate and approximate 32-bit RCAs and CLAs based on physical implementation using a 32/28nm CMOS process. Finally, Section V states the conclusions. Scope for further work exists to consider the application of the proposed approximate adders in the domain of digital signal processing.

P. Balasubramanian and C. Dang are with the School of Electrical and Electronic Engineering, Nanyang Technological University, 50 Nanyang Avenue, Singapore 639798, E-mails: balasubramanian@ntu.edu.sg; hcdang@ntu.edu.sg

D.L. Maskell is with the School of Computer Science and Engineering, Nanyang Technological University, 50 Nanyang Avenue, Singapore 639798, E-mail: asdouglas@ntu.edu.sg

K. Prasad is with the Department of Electrical and Electronic Engineering, Auckland University of Technology, Auckland 1142, New Zealand, E-mail: krishnamachar.prasad@aut.ac.nz

## II. ACCURATE RCA AND CLA

The circuit architectures of the 32-bit accurate RCA and CLA are shown in Figs 1a and 1b. A31 to A0 and B31 to B0 represent the 32-bit augend and addend inputs, with A31 and B31 being the most significant bits (MSBs), and A0 and B0 are the least significant bits (LSBs). C0 represents the carry input to the adder which may be prefixed to 0. SUM31 to SUM0 represents the sum output with SUM31 being the MSB and SUM0 is the LSB. C32 represents the carry output or overflow.

The 32-bit accurate RCA, shown in Fig 1a, is formed by a cascade of 32 full adders (FAs) which are labelled as FA1 to FA32. FA32 computes the MSB of the sum, and FA1 computes the LSB of the sum. In Fig 1b, a 32-bit accurate CLA is shown, which is realized using eight 4-bit sub-CLA modules which are labelled as CLA1 to CLA8. CLA8 computes the sum portion corresponding to the most significant nibble, and CLA1 computes the sum portion corresponding to the least significant nibble. The logic detail of the 4-bit CLA is depicted in Fig 1c, which consists of propagate-generate logic, a 4-bit carry lookahead generator (CLG), and the sum logic. In Fig 1c, P3 to P0 denote the propagate signals, and G3 to G0 represent the generate signals. C1 to C4 are the lookahead carry outputs with C4 being the carry lookahead input for the next stage CLA. The gate-level detail of the delay-optimized 4-bit CLG is shown in Fig 1d [19], which is synthesized using simple and complex logic gates. The 4-bit CLG shown in Fig 1d is said to be delay optimized since the carry input and the lookahead carry outputs are individually linked through a single complex gate viz. the AO21 gate. The generalized logic expressions corresponding to propagate and generate signals, the lookahead carry output, and the sum output are given below. In the equations, the symbols $\oplus$ and + imply logical EXOR and OR. The conjunction of two or more literals signifies the logical product.

$$P_i = A_i \oplus B_i \quad (1)$$

$$G_i = A_i B_i \quad (2)$$

$$C_{i+1} = G_i + P_i G_{i-1} + \ldots + P_i P_{i-1} \ldots C_i \quad (3)$$

$$SUM_i = P_i \oplus C_i \quad (4)$$

## III. APPROXIMATE RCAs AND CLAs

Approximations are introduced into certain least significant bit positions of an adder while retaining the more significant bit positions of the adder as accurate [11 – 14]. This is because the most significant adder bits have higher weights in terms of powers of 2 than the least significant adder bits. For example, the MSB of a 32-bit adder is associated with a weight of $2^{32}$ and the LSB of the adder is associated with a weight of just $2^0$. Hence, any error in the LSB positions may be tolerated and may not be visible while any error in the MSB positions may not be tolerated and may cause a visible deterioration in the output quality. Further, there exists a tradeoff between accuracy and energy efficiency of the approximate adders.

To realize an approximate adder, an *n*-bit adder may be split into two parts as the accurate adder part and the approximate adder part. Given this, addition can be performed in parallel in the accurate and approximate adder parts. If the accurate adder part comprises *m*-bits then the approximate adder part would comprise (*n* – *m*) bits. The number of bits to be allotted to the approximate adder part

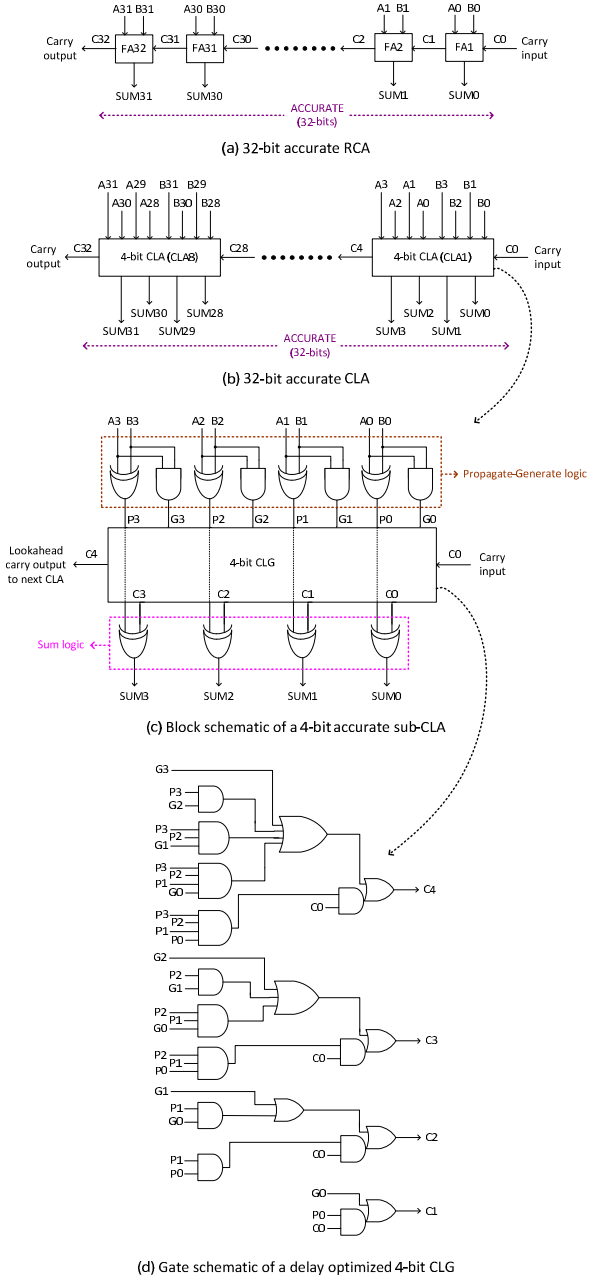

Fig. 1. (a) Accurate 32-bit RCA; (b) Accurate 32-bit CLA; (c) Internal block level detail of 4-bit CLA; (d) Gate-level detail of delay-optimized 4-bit CLG

may be custom-defined for a specific application based on the application's inherent error resiliency. Since the approximate adder part is not mandated to produce the correct sum, carry propagation may be eliminated in the approximate adder part. As a result, the individual sum outputs corresponding to the approximate adder part may be produced using just 2-input XOR or OR gates by involving only the respective augend and addend inputs and thereby neglecting any carry input. In other words, instead of the accurate sum production based on (4), an approximate sum output may be produced based on (5) or (6) which are given below. To realize the accurate adder part of an approximate adder, any adder topology such as RCA, CLA etc. may be used [20].

$SUM_i = A_i \oplus B_i$ (5)

$SUM_i = A_i + B_i$ (6)

Table I shows a comparison between the accurate sum output expressed by (4), and the two approximate sum outputs expressed by (5) and (6). It is noticed that when assuming a uniform inputs distribution, the use of either XOR or OR gates to synthesize an approximate sum bit results in an identical number of correct and incorrect outputs as shown in Table I. Hence the usage of 2-input XOR or OR gates may not make a difference to the sum output. However, since a 2-input XOR gate ($4.32\mu m^2$) is more expensive than a 2-input OR gate ($2.03\mu m^2$) by $1.1\times$ [21], 2-input OR gates may be used instead of 2-input XOR gates to produce the approximate sum bits based on (6). To increase the number of correct outputs in the approximate adder part, either approximate full adders [12 – 14] or approximate versions of accurate full adders [22] [23] may be used in place of the 2-input OR gates. Nevertheless, this will increase the logic and also increase the design metrics.

TABLE I
COMPARISON OF ACCURATE AND APPROXIMATE SUMS

| Primary inputs | | | Accurate Sum – Eqn. (4) | Approximate Sum – Eqn. (5) | Approximate Sum – Eqn. (6) |
|---|---|---|---|---|---|
| $A_i$ | $B_i$ | $C_i$ | | | |
| 0 | 0 | 0 | 0 | 0 (correct) | 0 (correct) |
| 0 | 0 | 1 | 1 | 0 (incorrect) | 0 (incorrect) |
| 0 | 1 | 0 | 1 | 1 (correct) | 1 (correct) |
| 0 | 1 | 1 | 0 | 1 (incorrect) | 1 (incorrect) |
| 1 | 0 | 0 | 1 | 1 (correct) | 1 (correct) |
| 1 | 0 | 1 | 0 | 1 (incorrect) | 1 (incorrect) |
| 1 | 1 | 0 | 0 | 0 (correct) | 1 (incorrect) |
| 1 | 1 | 1 | 1 | 0 (incorrect) | 1 (correct) |

Fig 2 shows four variants of approximate RCAs and CLAs. The carry input for the accurate adder parts of the approximate RCAs and CLAs are prefixed to 0. Approximation sizes ranging from 4 LSBs to 20 LSBs are considered through Figs 2a to 2e respectively. The number of bits allotted for the accurate and the approximate adder parts are highlighted in violet and green in Figs 2a to 2e. Approximate RCAs are realized by combining the accurate adder parts shown within the dotted red boxes in Figs 2a to 2e with the respective approximate adder parts. On the other hand, approximate CLAs are realized by combining the accurate adder parts shown within the dotted blue boxes in Figs 2a to 2e with the respective approximate adder parts.

## IV. RESULTS AND DISCUSSION

Accurate and approximate 32-bit RCAs and CLAs were realized in semi-custom ASIC design style using the elements of a 32/28nm CMOS digital standard cell library [21]. Only the minimum size cells were used to realize the accurate and approximate RCAs and CLAs to pave the way for a straightforward comparison of their design metrics post physical synthesis.

About 1000 random input vectors were identically supplied to the RCAs and CLAs through a test bench at time intervals of 4ns i.e. 250MHz. The functional simulations were performed using Synopsys VCS to verify the correctness of the respective adders synthesized. The .vcd files generated were subsequently used to estimate the average power dissipation. The time-based power analysis mode of Synopsys PrimeTime was invoked to accurately estimate the average power dissipation. The critical path delay and area occupancy were also estimated with default wire loads included. The design parameters estimated are given in Table II. The percentage values mentioned within brackets in Table II signify the corresponding percentage reductions in design metrics achieved for the approximate RCAs and CLAs compared to the design metrics of the accurate RCA and CLA respectively.

It can be noticed from Table II that as the approximation size increases from 4 LSBs to 20 LSBs, the design metrics of the approximate adders (RCAs or CLAs) decrease. To explain this phenomenon, let us consider the RCA architecture first for an illustration. The full adder of [21] occupies an area of $4.83\mu m^2$, while the 2-input OR gate of [21] occupies 58% less area of $2.03\mu m^2$. Hence, replacing the full adders by 2-input OR gates in the approximate adder part would decrease the areas of the approximate RCAs compared to the accurate RCA. Further, the savings in area would increase with increases in the degree of approximation. This savings in area would in turn translate into reductions in total power dissipation. An $n$-bit accurate RCA would experience $n$ full adder delays. In contrast, an $n$-bit approximate RCA with $m$ bits allotted to the accurate adder part would experience only $m$ full adder delays. As $m$ decreases in magnitude relative to $n$, the critical path delay of the approximate RCA would also decrease proportionately. Hence, approximate RCAs would feature reductions in power, delay, and area occupancy compared to the accurate RCA.

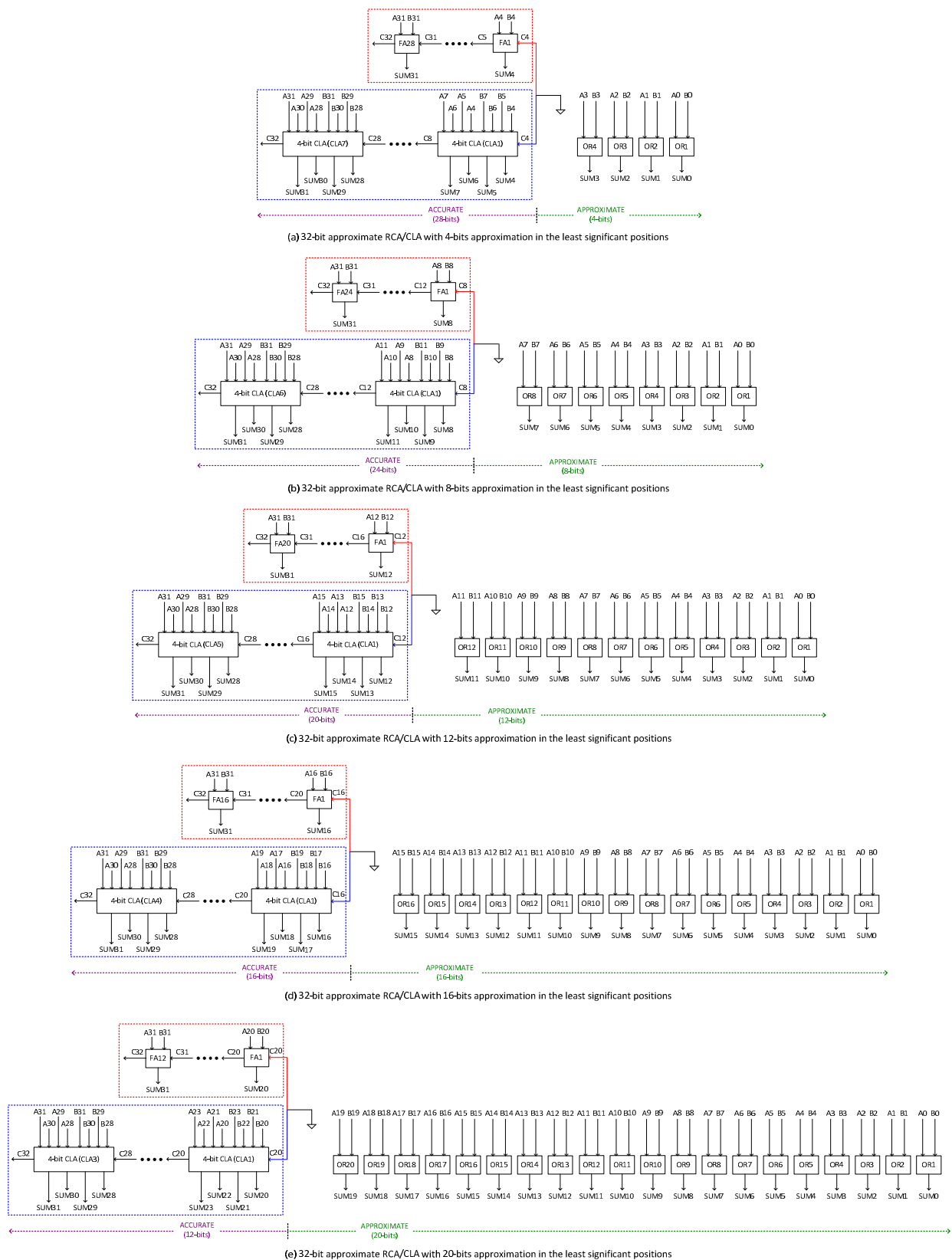

Fig. 2. 32-bits approximate RCA/CLA with: (a) 4-bits approximation, (b): 8-bits approximation, (c): 12-bits approximation, (d): 16-bits approximation, and (e): 20-bits approximation introduced in the less significant positions

TABLE II
AVERAGE POWER DISSIPATION, CRITICAL PATH DELAY, AND AREA
OF 32-BIT ACCURATE AND APPROXIMATE RCAS AND CLAS,
ESTIMATED USING A 32/28NM CMOS PROCESS

| Approximation size and Adder legend | Power (µW) | Delay (ns) | Area (µm$^2$) |
|---|---|---|---|
| *Accurate RCA* | | | |
| NA (RCA) | 35.18 | 3.35 | 154.52 |
| *Approximate RCAs* | | | |
| 4-bits (RCX1) | 32.28 (–8.2%) | 2.94 (–12.2%) | 143.34 (–7.2%) |
| 8-bits (RCX2) | 27.98 (–20.5%) | 2.53 (–24.5%) | 132.15 (–14.5%) |
| 12-bits (RCX3) | 23.69 (–32.7%) | 2.12 (–36.7%) | 120.97 (–21.7%) |
| 16-bits (RCX4) | 19.36 (–45%) | 1.70 (–49.3%) | 109.79 (–28.9%) |
| 20-bits (RCX5) | 16.41 (–53.4%) | 1.29 (–61.5%) | 98.61 (–36.2%) |
| *Accurate CLA* | | | |
| NA (CLA) | 49.05 | 1.13 | 646.54 |
| *Approximate CLAs* | | | |
| 4-bits (CLX1) | 44.41 (–9.5%) | 1.04 (–8%) | 573.86 (–11.2%) |
| 8-bits (CLX2) | 38.29 (–21.9%) | 0.95 (–15.9%) | 501.17 (–22.5%) |
| 12-bits (CLX3) | 32.10 (–34.6%) | 0.86 (–23.9%) | 428.49 (–33.7%) |
| 16-bits (CLX4) | 25.83 (–47.3%) | 0.77 (–31.9%) | 355.80 (–45%) |
| 20-bits (CLX5) | 21.05 (–57.1%) | 0.68 (–39.8%) | 283.12 (–56.2%) |

A similar explanation holds good for the approximate CLAs where the replacement of 4-bit sub-CLAs in the approximate adder part by 2-input OR gates would lead to enhanced savings in area, as evident from Table II. This is because the delay optimized 4-bit CLA shown in Fig 1c requires an area of 80.82µm$^2$. If this 4-bit CLA is to be replaced by four 2-input OR gates in the approximate adder part then an almost 9× reduction in area can be achieved. Again, the less area occupancy of approximate CLAs would imply lesser power dissipation for them compared to the accurate CLA. The critical path delay of an *n*-bit accurate CLA, as shown in Fig 1b, is governed by a factor of (*n*/4), since the sub-CLA module is of size 4-bits. In contrast, the critical path delay of an approximate CLA would be governed by a factor of (*m*/4). As *m* becomes lesser than *n*, the critical path delay of an approximate CLA would decrease proportionately. Therefore, approximate CLAs would be better optimized than the accurate CLA.

The power-delay product (PDP) is a standard metric [24] that is used to evaluate the low power attribute of a digital circuit or system. Since total power dissipation and maximum propagation delay are desirable to be minimized, the lesser the PDP of a digital circuit or system, the better optimized is its design. The PDPs of accurate and approximate 32-bit RCAs and CLAs were calculated based on Table II and are plotted in Fig 3 for comparison. Since the power and delay of approximate RCAs and CLAs are reduced compared to those of the accurate RCA and CLA, as evident from Table II, the PDP metrics of the former are lesser compared to the PDP metrics of the latter as seen in Fig 3. The averaged PDP of approximate RCAs is less than the PDP of the accurate RCA by 54.2%. Also, the averaged PDP of approximate CLAs is less than the PDP of the accurate CLA by 47.9%.

Referring to Table II, it can be noted that the mean of the power dissipations of approximate CLAs is 32.34µW, and the mean of the power dissipations of approximate RCAs is 23.94µW which signifies a 26% decrease. This decrease is expected since the approximate RCAs occupy 71.8% less area on average compared to the approximate CLAs. On the other hand, the mean of the critical path delays of the approximate RCAs is 2.12ns, and the mean of the critical path delays of the approximate CLAs is 0.86ns which signifies a 59.4% decrease. This decrease is mainly because of the logarithmic delay magnitude achievable in the case of the CLA architecture compared to the linear delay magnitude defining the RCA architecture. Thus the mean decrease in total power dissipation achieved by the approximate RCAs over the approximate CLAs is offset by the greater decrease in the mean critical path delay achieved by the latter over the former. Therefore, overall, the approximate CLAs report lesser PDP than the approximate RCAs.

V. CONCLUSIONS

This paper has presented approximate RCAs and CLAs and has made a comparison with the accurate RCA and CLA. A 32-bit addition operation was considered as the case study, and the accurate and approximate RCAs and CLAs were realized using a 32/28nm CMOS technology. The simulation results show that the approximate RCAs and CLAs report significant reductions in PDP and area compared to the accurate RCA and CLA respectively. However, the optimizations in design metrics are achieved at the expense of a trade-off in terms of the accuracy of results. Based on the acceptable accuracy of results, the degree of approximation may be predetermined for a specific application. Our analysis shows that for the same range of approximation sizes considered viz. 4 LSBs to 20 LSBs, the approximate CLAs achieve a 46.5% reduction in PDP compared to the approximate RCAs. Given this observation, it is noted that the approximate CLA architecture is preferable over the approximate RCA architecture for a low power realization of approximate computer arithmetic. Also, with respect to high-speed, the approximate CLA architecture is preferable than the approximate RCA architecture. The main advantage of the approximate RCAs is their less area occupancy.

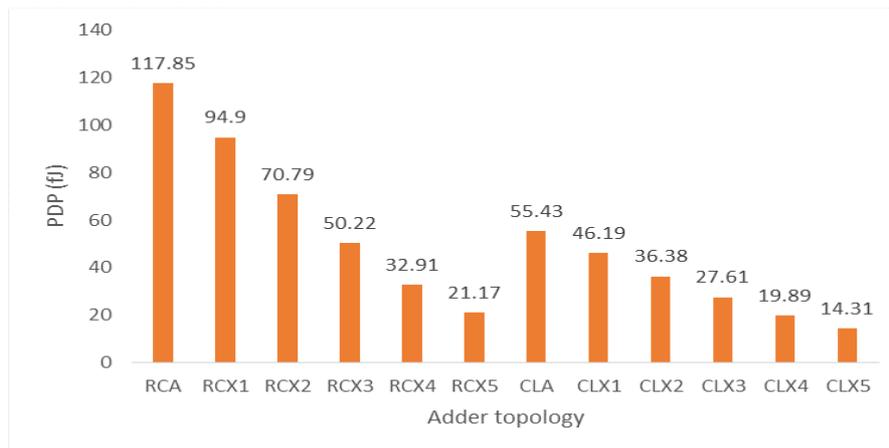

Fig. 3. PDP values of accurate and approximate 32-bit RCAs and CLAs (refer column 1 of Table II for the legends specified within brackets, which are used to represent the respective adder topologies in the X-axis)